\documentclass[11pt,a4paper]{article} 
\usepackage{latexsym,amsmath,amssymb}
\usepackage{graphicx}
\usepackage{slashed}  
\usepackage{bm}
\usepackage{epsfig}
\usepackage{dcolumn}
\usepackage{subcaption}

\usepackage{geometry}
\usepackage{setspace}
\newcommand{\De}{\Delta}
\newcommand{\Om}{\Omega}
\newcommand{\fr}{\frac}
\newcommand{\Ra}{\Rightarrow}

\newcommand{\txt}{\textnormal}
\newcommand{\ze}{\zeta}

\newcommand{\beq}{\begin{equation}}
\newcommand{\eeq}{\end{equation}}
\newcommand{\bea}{\begin{eqnarray}}
\newcommand{\eea}{\end{eqnarray}}
\newcommand{\bef}{\begin{figure}}
\newcommand{\eef}{\end{figure}}
\newcommand{\al}{\alpha}
 
\newcommand{\e}{\epsilon}

\newcommand{\Gam}{\Gamma}

\newcommand{\Ome}{\Omega}

\newcommand{\lmk}{\left(}
\newcommand{\rmk}{\right)}

\newcommand{\lle}{\left<}
\newcommand{\rgr}{\right>}

\newcommand{\pr}{^{\prime}}

\def\baq{\begin{eqnarray}}
\def\eaq{\end{eqnarray}}

\def\k{{\bf k}}

\def\pa{\partial}

\def\lesssim{~\mbox{\raisebox{-.6ex}{$\stackrel{<}{\sim}$}}~}

\def\ltap{\ \raise.3ex\hbox{$<$\kern-.75em\lower1ex\hbox{$\sim$}}\ }
\def\gtap{\ \raise.3ex\hbox{$>$\kern-.75em\lower1ex\hbox{$\sim$}}\ }
\def\gl{\ \raise.5ex\hbox{$>$}\kern-.8em\lower.5ex\hbox{$<$}\ }
\def\roughly#1{\raise.3ex\hbox{$#1$\kern-.75em\lower1ex\hbox{$\sim$}}}

\begin{document}
\thispagestyle{empty}
\begin{titlepage}
\nopagebreak

\title{\bf\Large Inflationary Magnetogenesis  without the Strong Coupling Problem II: \\
Constraints from CMB anisotropies and B-modes}
%\begin{center}
%\end{center}

\vfill
\author{Ricardo J. Z. Ferreira\footnote{ferreira@cp3.dias.sdu.dk},~ Rajeev Kumar Jain\footnote{jain@cp3.dias.sdu.dk} ~and Martin S. Sloth\footnote{sloth@cp3.dias.sdu.dk}
}
\date{ }

%\end{center}

\maketitle

\begin{center}

%\vskip -1.2cm

%\vskip 0.5cm
{\it  CP$^3$-Origins, Center for Cosmology and Particle Physics Phenomenology}\\

{\it  University of Southern Denmark, Campusvej 55, 5230 Odense M, Denmark}

%\vskip -1.2cm

\end{center}
\vfill
\begin{abstract}
Recent observational claims of magnetic fields stronger than $10^{-16}$ G in the extragalactic medium motivate a new look for their origin in the inflationary magnetogenesis models. In this work we shall review the constraints on the simplest gauge invariant model $f^2(\phi)F_{\mu \nu}F^{\mu \nu}$ of inflationary magnetogenesis, and show that in the optimal region of parameter space the anisotropic constraints coming from the induced bispectrum, due to the generated electromagnetic fields, yield the strongest constraints. In this model, only a very fine tuned scenario at an energy scale of inflation as low as $10^{-2}$ GeV can explain the observations of void magnetic fields. These findings are consistent with the recently derived upper bound on the inflationary energy scale. However, if the detection of primordial tensor modes by BICEP2 is confirmed, the possibility of low scale inflation is excluded. Assuming the validity of the BICEP2 claim of a tensor-to-scalar ratio $r=0.2^{+0.07}_{-0.05}$, we provide the updated constraints on this model of inflationary magnetogenesis. On the Mpc scale, we find that the maximal allowed magnetic field strength from inflation is less than $10^{-30}$ G.

 \end{abstract}
 %\vskip.4in
\noindent
DNRF90
\hfill \\
\vfill
\end{titlepage}

\section{Introduction}

Large scale coherent magnetic fields are omnipresent across the entire universe. While their presence in cosmic structures e.g. stars, galaxies and clusters of galaxies has been verified by different astronomical observations, their true origin has not yet been entirely understood (see, for instance, \cite{Ryu:2011hu, Widrow:2011hs,Durrer:2013pga} for recent reviews). It is widely believed that subsequent enhancement of pre-existing seed fields due to the galactic dynamo mechanism \cite{Brandenburg:2004jv} could lead to such magnetic fields although the strength of seed fields must be larger than $10^{-20}-10^{-30}$ G \cite{Turner:1987bw, Davis:1999bt}. Recent indirect observations of femto-Gauss magnetic fields in voids with a coherence length larger than the Mpc scale have further intensified the search for their origin \cite{Neronov:1900zz,Tavecchio:2010ja,Tavecchio:2010mk,Taylor:2011bn}. The large coherence length of void magnetic fields makes them difficult to be produced in the late universe and hints towards their origin during the inflationary epoch in the early universe.

Among various possibilities, inflationary magnetogenesis has been considered a plausible mechanism for the generation of such cosmic magnetic fields. One of the simplest, gauge invariant and well studied model of inflationary magnetogenesis is described by the Lagrangian \cite{Turner:1987bw,Ratra:1991bn}
\beq
{\cal{L}}_{\rm EM}=-\frac{1}{4}f^2(\phi)F_{\mu \nu}F^{\mu \nu},
\label{eq:Lag}
\eeq
where $F_{\mu \nu}$ is the electromagnetic (EM) field tensor and is defined as $F_{\mu \nu} \equiv \pa_\mu A_\nu-\pa_\nu A_\mu$. In this model the conformal invariance of the $U(1)$ gauge field $A_\mu$ is broken by a time dependent function $f$ of a dynamical scalar field $\phi$ thereby generating an effective coupling constant $e_{\rm eff}=e/f$. Although this model has been greatly studied, there exists certain issues regarding its ability to explain the observations consistently. In \cite{Demozzi:2009fu}, it was discussed that this model always suffers from one of two main problems: backreaction due to the energy density of generated EM fields on the inflationary dynamics or strongly coupled regimes at the onset of inflation where the theory loses its predictability. The problem of strong coupling can be avoided by dropping the requirement of gauge invariance at high energies, as suggested in \cite{Bonvin:2011dt}, but we are not aware of any explicit models that can achieve this and restore gauge invariance at the end of inflation as required. Instead, in \cite{Ferreira:2013sqa}, this no-go theorem was circumvented without breaking gauge invariance by lowering significantly the energy scale of inflation wherein femto-Gauss magnetic fields could be achieved for a TeV scale inflation.

Apart from the aforementioned two problems, there exists yet another problem, the anisotropies from the curvature perturbations induced by these EM fields generated during inflation may become too large \cite{Fujita:2014sna,Fujita:2012rb,Fujita:2013qxa,Nurmi:2013gpa}. In \cite{Fujita:2013qxa}, the primordial magnetic field strength was strongly constrained in the model described in eq. (\ref{eq:Lag}) by requiring that the induced perturbation spectrum at CMB scales must be in agreement with the recent Planck observations \cite{Ade:2013uln, Ade:2013ydc}. It was also concluded that the backreaction constraint is generically stronger than the constraint from curvature perturbations when the duration of inflation is much longer than the minimum required to solve the horizon problem. As we will see, the role of the two constraints are interchanged when the duration of inflation is close to the minimal required to solve the horizon problem. This should be compared with the conservative upper bound obtained for the magnetic field strength today in \cite{Fujita:2014sna}\footnote{The expression for the minimum amount of inflation appearing in the Planck paper \cite{Ade:2013uln}, also used by the authors of \cite{Fujita:2014sna}, carries a misprint \cite{pp}. The equation in \cite{Ade:2013uln} gives, for instantaneous reheating and neglecting slow-roll terms, $N_{\rm min}\simeq 71.5+1/2 \log(H/M_p)$ where $M_p$ is the reduced Planck mass, defined as $M_p^2 \equiv 1/{8 \pi G}$. However, in \cite{Ferreira:2013sqa} the same quantity was computed yielding $N_{\rm min}\simeq 66.9+1/2 \log(H/M_p)$ which is in agreement with \cite{Liddle:2003as}.} 
\beq
\rho^{1/4}_{\rm inf}\ <\ 29.3\, \txt{GeV} \lmk \frac{k}{1\, \text{Mpc}^{-1}} \rmk^{5/4} \lmk \frac{B_{0}}{10^{-15}\, \txt{G}}\rmk^{-1},\quad k>1\, \text{Mpc}^{-1},
\label{eq:bound} 
\eeq
where $\rho_{\rm inf}$ is the energy density during inflation and $B_{0}$ is the present day magnetic field strength.
This upper bound was derived under the requirement of gauge invariance and non-strongly coupled regimes and is valid in the region where the electric field energy dominates over the magnetic field.

In this paper, we shall review these constraints in the specific case of the coupling $f^2(\phi)F_{\mu \nu}F^{\mu \nu}$. We discuss backreaction constraint as well as anisotropic constraints both from the induced power spectrum and non-Gaussianities. We shall show that in the case where inflation lasts an amount of e-folds close to the minimum required to solve the horizon problem, the magnetic field strength today is constrained to be $10^{-15}$ G at $10$ MeV which is approximately $3$ orders of magnitude lower than the conservative upper bound derived in \cite{Fujita:2014sna}.  Finally, in view of the recent observations of primordial tensor modes through the B-mode polarization of the CMB \cite{Ade:2014xna}, we update the resulting constraints on inflationary magnetogenesis given that the referred observations, if correct, fix the energy scale of inflation to be $\rho^{1/4}_{\rm inf} \simeq 10^{16}$ GeV which has strong consequences on the results discussed in the previous sections.

%%%%%%%%%%%%%%%%%%%%%%%%%%%%%%%%%%%%%%%%%%%%%%%%%%%%%%%%%%%%%%%%%%%%%%%%
\section{A simple model: $f^2(\phi)F_{\mu \nu}F^{\mu \nu}$}

From a phenomenological point of view, we do not either need to specify the dynamical scalar field $\phi$ nor the underlying physics leading to the $f(\phi)$ coupling in the Lagrangian in eq. (\ref{eq:Lag}). Instead, the coupling function can be parametrized as $f(\phi)\propto a^{\al}$ where $a$ is the scale factor and the only requirement is to recover the standard electromagnetism at the end of inflation\footnote{A model where this condition is relaxed and the coupling is also allowed to have non-trivial time-dependence after inflation was very recently proposed in \cite{Kobayashi:2014sga}. In this case it is suggested that appreciable magnetic fields can be produced after inflation, but before reheating. In the present work we have however restricted ourselves to consider magnetic fields generated during inflation.} i.e. $f(a_{\rm end}) \to 1$.

It is well known that for $\al>0$, strongly coupled regimes during inflation are unavoidable \cite{Demozzi:2009fu,Ferreira:2013sqa} and therefore, $\al<0$ is required to have a consistent theory. In this regime, the strength of magnetic fields today can be written as \cite{Ferreira:2013sqa}

\beq
B_k(\al,H)=\fr{\Gam(-\al-1/2)}{2^{3/2+\al}\pi^{3/2}}\, H^2 \lmk R\, \Om_r^{1/4} \rmk^{-(1+\al)} \lmk \fr{H_0}{H} \rmk ^{\frac{1}{2}(5+\al)} \lmk \frac{k}{a_0 H_0} \rmk ^{3+\al},
\label{eq:B}
\eeq
where $\Om_r \sim 5.4 \times 10^{-5} $ is the present radiation density parameter, $H_0 \sim 6 \times10^{-61} M_p$ is the Hubble constant today \cite{Ade:2013zuv}, $H$ is the Hubble parameter during inflation and $R$ is the reheating parameter. Instantaneous reheating corresponds to $R=1$. A naive look at eq. ({\ref{eq:B}}) would immediately indicate that it is possible to generate sufficiently large magnetic fields in this model. However, a set of consistency checks must be done before drawing such a conclusion. 

%%%%%%%%%%%%%%%%%%%%%%%%%%%%%%%%%%%%%%%%%%%%%%%%%%%%%%%%%%%%%%%%%%%%%%%%
\subsection{Constraints from backreaction}

The backreaction constraint comes from requiring that the energy density of the produced EM fields does not backreact on the background inflationary dynamics. For $\al<-2$, a regime for which there are no strongly coupled regimes and the magnetic field is effectively excited, the EM energy density $\rho_{\rm em}$ is mainly stored in the electric field and is maximal at the end of inflation yielding \cite{Ferreira:2013sqa, Fujita:2013qxa}
\beq
\rho_{\rm em} \simeq d_{\al}\, H^4\, e^{-(2\al+4)(N_{\rm tot} - N_{\rm b})}, \quad d_{\al} \equiv -\frac{\Gam^2(1/2-\al)}{2^{2\al+2}\,\pi^{3}(2\al+4)}
\eeq
where $N_{\rm b}$ is the number of e-folds after the beginning of inflation at which the conformal coupling is broken and we have used some useful identities for the Gamma function in order to simplify our expression. We have also assumed that inflation lasts for $N_{\rm tot}=N_{\rm min}+\De N$ e-folds with 
\beq
N_{\rm min}= \ln(R)+\frac{1}{2}\ln\left(\frac{H}{H_{0}}\right)+\frac{1}{4}\ln(\Omega_{r}),
\eeq
the minimum amount of inflation required to solve the horizon problem \cite{Ferreira:2013sqa, Liddle:2003as}. Therefore, in the case of instantaneous reheating\footnote{In this paper we restrict ourselves to the case of instantaneous reheating. One possible extension of this scenario would be to consider the existence of a prolongated reheating stage \cite{Ferreira:2013sqa, Demozzi:2012wh, Ringeval:2013hfa, Martin:2007ue} where the background has an equation of state parameter $w$ different than radiation i.e $w \neq 1/3$. Although \cite{Demozzi:2012wh, Ringeval:2013hfa} did not consider the strong coupling problem but approached the problem rather model independently. } requiring that there is no backreaction translates into the condition $\rho_{\rm em}<\rho_{\rm inf}$ which can be solved for $H$ leading to
\beq
\lmk \frac{H}{H_0} \rmk^{-\al} < \frac{3\,\Ome_r^{(\al/2+1)}}{d_\al}\lmk \frac{M_p}{H_0} \rmk^2e^{(2\al+4)(\De N-N_{\rm b})}.
\label{eq:BC}
\eeq
Disregarding the term in $d_{\al}$, we can obtain an analytical inequality for $\al$ as it was done in \cite{Ferreira:2013sqa},
\beq
\al \gtrsim-2+\frac{\ln \lmk \frac{H}{M_p} \rmk }{\fr{1}{2}\ln \lmk \frac{H}{H_0}\Ome_r^{1/2} \rmk +\Delta N - N_b}.
\label{eq:BCal}
\eeq
By inserting this minimal allowed value of $\al$ in eq. (\ref{eq:B}), we obtain the maximal value of the magnetic field strength today, allowed by backreaction, as a function of $H$ at a given length scale. 

%%%%%%%%%%%%%%%%%%%%%%%%%%%%%%%%%%%%%%%%%%%%%%%%%%%%%%%%%%%%%%%%%%%%%%%%
\subsection{Anisotropy constraints}

Another important feature of generating EM fields during inflation is that they lead to non-adiabatic pressure perturbations which can source the adiabatic perturbations at super horizon scales. This additional contribution leads to distinct features in the CMB both at the level of the power spectrum and non-Gaussianities, which could be large enough for detection. In the presence of non-adiabatic perturbations, the time evolution of the curvature perturbations in the super horizon regime is given by \cite{Nurmi:2013gpa,Wands:2000dp}
\beq
\dot{\ze}=-\frac{H}{\rho_t+p_t}\delta p_{\rm nad},
\label{eq:pert_generic}
\eeq
where $\rho_t$ is the total energy density, $p_t$ is the total pressure and $\delta p_{\rm nad} \equiv \delta p_t- \frac{\dot{p_t}}{\dot{\rho_t}} \delta \rho_t$ is the so-called non-adiabatic pressure. During inflation $p_t\simeq(-1+2/3 \epsilon)\rho_t$, while in the presence of EM fields, $\delta p_{\rm nad} \simeq \frac{4}{3}\delta \rho_{\rm em}$ and therefore, they contribute to the curvature perturbation $\ze$ as
\beq
\ze_{\rm em}(\k,t\pr)=-\frac{2H}{\e\, \rho_t} \int_{t_{\rm exit}}^{t \pr} dt\, \delta \rho_{\rm em} (\k,t)
\eeq
where $\e$ is the first slow-roll parameter and $t_{\rm exit}$ is the time of horizon crossing of the mode.
If the observed perturbations are not sourced by the inflaton, the strongest requirement one should impose is that the power spectrum generated by $\ze_{\rm em}$ at CMB scales is smaller than the total observed spectrum ${\cal{P}}^{\rm obs}_{\ze}\simeq 2.2\times10^{-9}$. In fact, $\ze_{\rm em}$ can contribute to the power spectrum in two different ways, through the term ${\cal{P}}_{\ze_{\rm em}}\propto \lle \ze_{\rm em} \ze_{\rm em}\rgr$  \cite{Fujita:2013qxa, Nurmi:2013gpa}, but also through the cross-correlations with the curvature perturbation generated by the auxiliary field appearing in the coupling function\footnote{In the $f^2(\phi)F_{\mu \nu}F^{\mu \nu}$ model, $b_{NL}=n_B-4$ is determined, in the squeezed limit, by the magnetic consistency relation \cite{Jain:2012ga,Jain:2012vm}.  The $b_{NL}$ parameter can also be probed by the consistency relations for magnetic fields in large scale structure \cite{Berger:2014wta}.} ${\cal{P}}^{b_{NL}}_{\ze_{\rm em}}\propto\lle \ze_{\phi}\ze_{\rm em}\rgr$ \cite{Nurmi:2013gpa}, at first order in the in-in formalism.
In this specific model, and for $\al<-2$, ${\cal{P}}_{\ze_{\rm em}}$ is approximately given, at the end of inflation, by 
\beq
{\cal{P}}_{\ze_{em}}(k)\simeq -\frac{16}{3(2\al+4)} \lmk \frac{H^2 d_{\al} }{3 \e M_p^2} \rmk^2  \lmk e^{-(2\al+4)(N_{\rm tot}-N_{\rm b}-N_{\rm k})}-1\rmk \lmk e^{-(2\al+4) N_{\rm k}} - 1\rmk ^2 ,
\label{eq:PS}
\eeq
where $N_{\rm k}$ is the e-fold of horizon exit of the mode $k$ counting from the end of inflation backwards in time. For $\al<-2$ the other contribution, ${\cal{P}}^{b_{NL}}_{\ze_{\rm em}}$,  is approximately given by
\beq
{\cal{P}}^{b_{NL}}_{\ze_{em}}(k)\simeq -\frac{b_{NL} {\cal{P}}_{\ze}(k)}{(2\al+4)} \lmk \frac{H^2 d_{\al} }{6 \e M_p^2}  \rmk  e^{-(2\al+4)(N_{\rm tot}-N_{\rm b})}.
\label{eq:PSbnl}
\eeq
One can easily check that for a reasonable value of $b_{NL}$, the former term ${\cal{P}}_{\ze_{\rm em}}(k)$ will give a much stronger constraint on the energy scale of inflation and therefore, we will not consider the constraint coming from the later term. 

If the conformal breaking occurs at the beginning of inflation, the largest value of ${\cal{P}}_{\ze_{\rm em}}(k)$ at CMB scales occurs for the horizon mode which leads to the strongest constraint on $H$. Therefore, requiring ${\cal{P}}_{\ze_{\rm em}}^{\rm max}< {\cal{P}}^{\rm obs}_{\ze}$, at CMB scales, implies that
\beq
\lmk \frac{H}{H_0} \rmk ^{-2\al} \lesssim  -\frac{3}{16}(2\al+4)  \lmk \frac{3  \e M_p^2}{d_{\al}H_0^2} \rmk ^2 {\cal{P}}^{\rm obs}_{\ze}\, \Ome_r^{\al+2}\, e^{-2(2\al+4)N_{\rm b}}\lmk e^{-(2\al+4)\De}-1 \rmk ^{-1},
\label{eq:PSC}
\eeq
where $\De=\De N$ if the conformal breaking occurs before the CMB modes left the horizon and $\De=0.1$ otherwise\footnote{If one considers exactly the minimum amount of inflation the power spectrum is maximum not at the horizon scale but about 0.1 e-folds later. We will, however, not consider this case.
In the case where the conformal breaking occurs when a mode smaller than the horizon scale becomes super horizon, the power spectrum is maximized again for a mode leaving the horizon about 0.1 e-folds later.}. As done in eq. (\ref{eq:BCal}), in order to solve the inequality for $\al$, we neglect the linear terms and approximate the last term of the above equation by $e^{(2\al+4) \De}$.  By doing so, we arrive at the expression
\beq
\al \gtrsim\frac{\ln \lmk \frac{3}{16} \lmk \frac{3 M_p^2 \e}{H_0^2} \rmk ^2{\cal{P}}^{\rm obs}_{\ze} \rmk +2 \ln \lmk \Ome_r \rmk +4 \De -8 N_b}{-2\ln \lmk \frac{H}{H_0} \Ome_r^{1/2} \rmk -2 \Delta + 4N_b},
\label{eq:PCal}
\eeq
which can now be inserted in eq. (\ref{eq:B}) to find the maximal value of $B$ in terms of the energy scale of inflation.
The same reasoning can be applied to the induced bispectrum. In that case the non-linearity parameter $f^{\rm loc}_{NL}$ generated by the correlator $\lle \ze_{\rm em} \ze_{\rm em} \ze_{\rm em}\rgr$ yields \cite{Fujita:2013qxa}
\beq
f_{NL}^{\rm em}=-\frac{20}{27\,(2\al+4)} \lmk {\cal{P}}^{\rm obs}_{\ze} \rmk^{-2} \lmk \frac{2d_{\al} H^2}{3 \e M_p^2} \rmk ^3 \lmk e^{-(2\al+4)(N_{\rm min}-N_{\rm b}-N_{\rm k})}-1 \rmk e^{-3(2\al+4)N_{\rm k}}.
\label{eq:fNL}
\eeq
In \cite{Nurmi:2013gpa} other possible contributions to the non-linearity parameter $f_{NL}$ associated with the correlators $\lle \ze_{} \ze_{} \ze_{\rm em}\rgr$ and $\lle \ze_{} \ze_{\rm em} \ze_{\rm em}\rgr$ were computed. However, we have verified that for $\al<-2$ these two contributions are sub-dominant in the squeezed limit.
From the recent Planck results \cite{Ade:2013ydc}, the constraint $f^{\rm loc}_{NL}<8.5\,(68\%$ CL)\footnote{The results do not change significantly if one considers the constraint on $f^{\rm loc}_{NL}$ at $95\%$ CL instead.} translates the above equation into the following inequality
\beq
\lmk \frac{H}{H_0} \rmk ^{-3\al} \lesssim -\frac{27}{20} (2\al+4)f^{\rm loc}_{NL}\lmk \frac{3 M_p^2 \e}{2d_{\al}H_0^2} \rmk^3  \lmk {\cal{P}}^{\rm obs}_{\ze} \rmk^2  \Ome_r^{3/2(\al+2)} e^{-3(2\al+4)N_{\rm b}}\lmk e^{-(2\al+4)\De}-1 \rmk ^{-1}.
\label{eq:fNLC}
\eeq
We can again proceed in the same way to arrive at an inequality for $\al$ as
\beq
\al \gtrsim \frac{\ln \lmk \frac{27}{20}f^{\rm loc}_{NL}  \lmk {\cal{P}}^{\rm obs}_{\ze} \rmk^2 \lmk \frac{3 M_p^2 \e}{2H_0^2} \rmk ^3\rmk +3 \ln \lmk \Ome_r \rmk +4 \De -12 N_b}{-3\ln \lmk \frac{H}{H_0}\Ome_r^{1/2} \rmk -2 \Delta + 6N_b}.
\label{eq:fNLCal}
\eeq
The minimum value for $\al$ gives the maximal magnetic field for a given $H$. Given that the bispectrum leads to a stronger constraint than the power spectrum, it would be natural to proceed and compute the trispectrum. However, we have verified that using the recent Planck results $\tau^{\rm loc}_{NL}<2800\,(95\% CL)$ the trispectrum constraint is very similar to the one from the bispectrum.

\bef
\begin{centering}
\includegraphics[width=8.4cm,height=6cm]{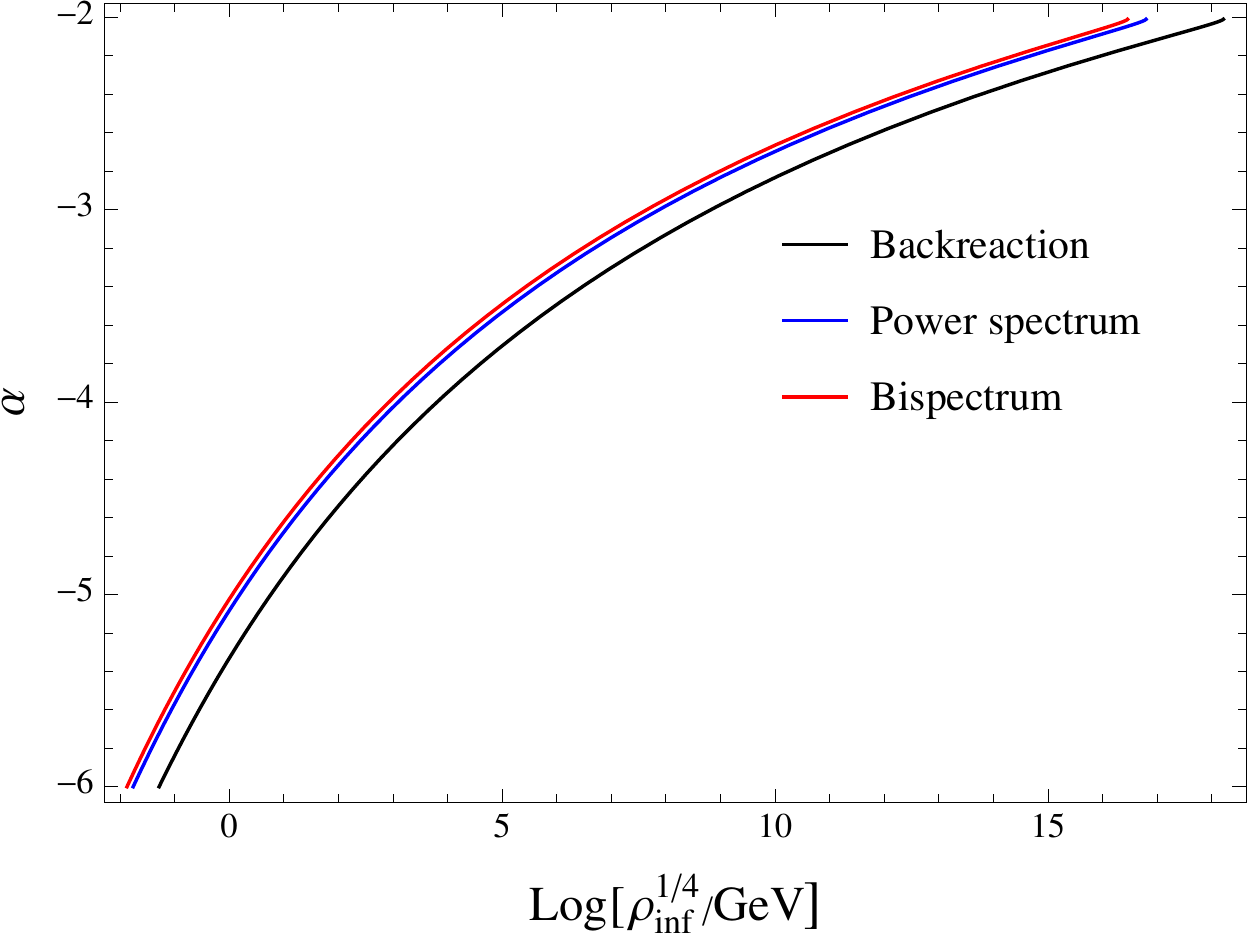}
\includegraphics[width=8.4cm,height=6cm]{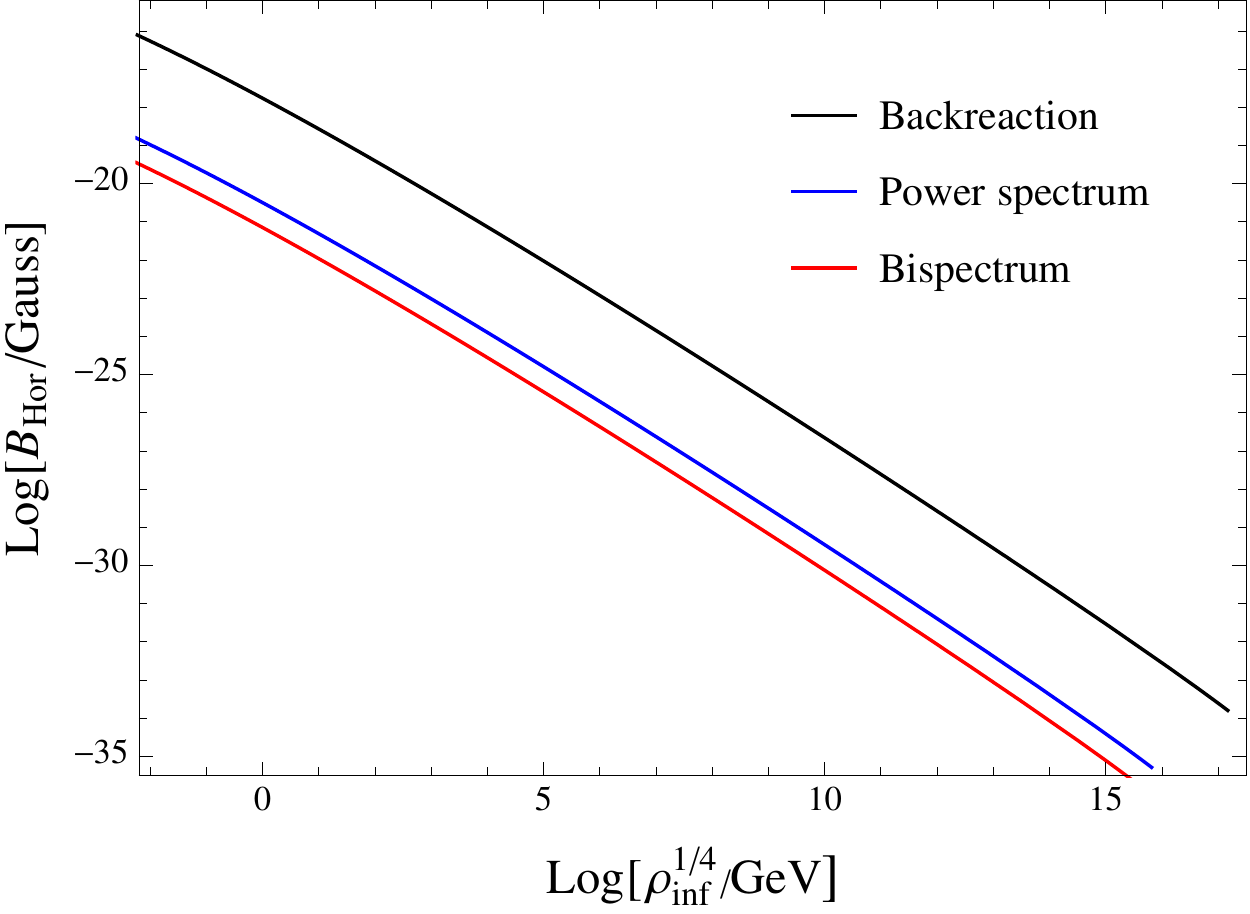}
%\includegraphics[scale=0.392]{B(rho)Standard_OurExp_Hor}
%\par
\end{centering}
\caption{Comparison between the constraints on the energy scale of inflation $\rho_{\rm inf}^{1/4}$ coming from backreaction, power spectrum and bispectrum (left) and the associated maximal magnetic fields at the horizon scale (right). The conformal coupling is broken since the beginning of inflation and we have assumed that inflation lasts 1 e-folds more than the minimum required.}
\label{fig:HC}
\eef

\subsection{Results}

In the previous section, we derived constraints on $H$ from backreaction, power spectrum and bispectrum in eqs. (\ref{eq:BC}), (\ref{eq:PSC}) and (\ref{eq:fNLC}), respectively. We can now compare them and obtain the maximal strengths allowed for magnetic fields. For clarity and in order to make a direct comparison with the literature, we translate the constraints on $H$ to constraints on the energy scale of inflation by using $\rho_{\rm inf}=3H^2M_p^2$. On the other hand, we also know that the energy scale of inflation is bounded between Big-Bang Nucleosynthesis $\rho^{1/4}_{\rm nucl} \simeq 10$ MeV and $\rho^{1/4} \simeq 10^{16}$ GeV and we will therefore work in this regime.  

In Fig. (\ref{fig:HC}) we plot the three different constraints on $\rho_{\rm inf}^{1/4}$ and the corresponding maximal magnetic field at the horizon scale. We find that there exists a hierarchy of constraints, namely, the constraint from the bispectrum is the strongest while the one coming from the power spectrum is quite close and the backreaction constraint is the weakest. This distinctive hierarchy is preserved in all the scenarios considered in this paper. Nevertheless, the role of the backreaction constraint increases with the total number of e-folds of inflation. In fact, if inflation lasted approximately $5$ e-folds more than $N_{\rm min}$, all the constraints are comparable to each other. If we consider even more e-folds of inflation, the hierarchy is reversed, as was also verified in \cite{Fujita:2013qxa}. However, we will stick to the hierarchy stated earlier where the strongest constraint comes from the bispectrum because all the other scenarios would lead to even weaker magnetic fields today. 
It is evident from Fig. (\ref{fig:HC}) that in the standard scenario wherein the conformal coupling is broken at the beginning of inflation and inflation lasts just $1$ e-fold more than the minimum required, the magnetic fields are maximal at the lowest energy scale, namely, for the horizon scale, the maximal value is $B \sim 10^{-19}$ G at $\rho_{\rm inf}^{1/4} \simeq 10^{-2}$ GeV and only $\sim 10^{-35}$ G at $\rho^{1/4}_{\rm inf} \simeq 10^{15}$ GeV. 

\bef
\begin{centering}
%\includegraphics[scale=0.4]{B(rho)StandardVsLate_OurExp}
%$\includegraphics[scale=0.4]{B(rho)_CompBound_Ourexp}
\includegraphics[width=8.4cm,height=6cm]{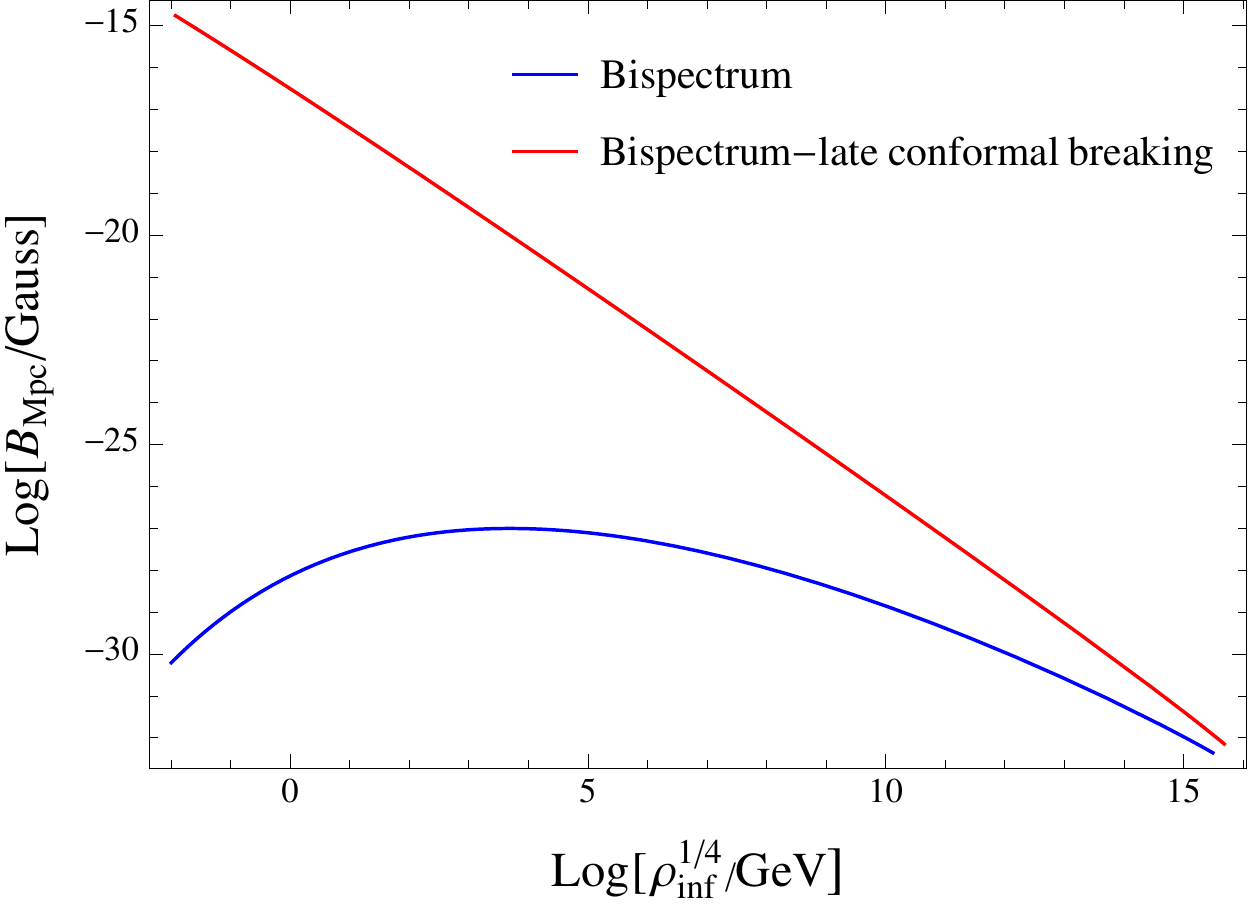}
\includegraphics[width=8.4cm,height=6cm]{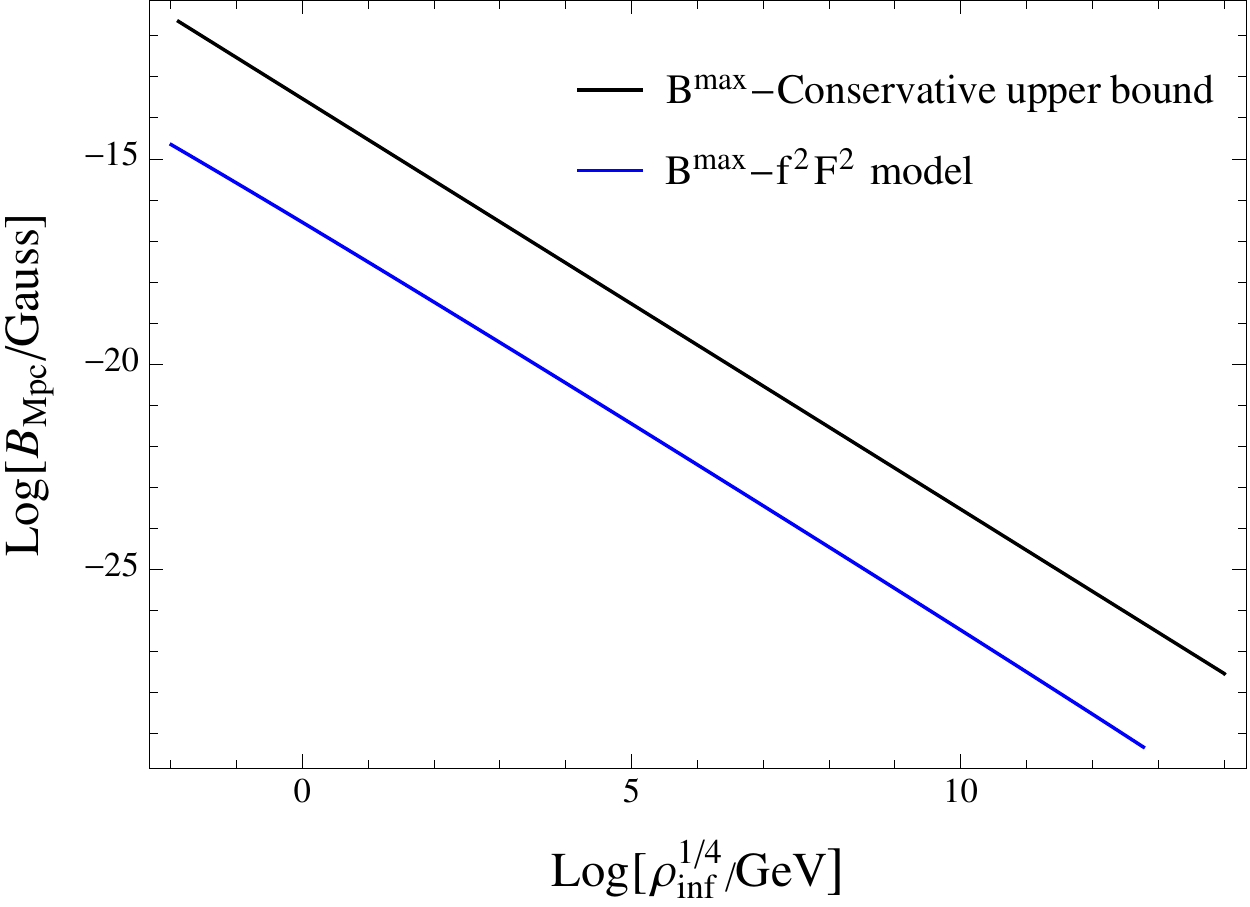}
%\par
\end{centering}
\caption{On the left, we compare the maximal magnetic fields at the Mpc scale allowed by the bispectrum constraint in the two scenarios where the conformal coupling is broken in the beginning of inflation and when it is only broken when the Mpc scale leaves the horizon. On the right side, we compare the maximal magnetic fields allowed in the $f^2(\phi) F^2$ model by anisotropic constraints with the maximal magnetic fields allowed by the conservative upper bound in eq. (\ref{eq:bound}), both at the Mpc scale.}
\label{fig:BfNL}
\eef

Given that the bispectrum leads to the strongest constraint, in what follows, we just plot the resulting constraints on magnetic fields. In Fig. (\ref{fig:BfNL}), we compare the maximal magnetic field at the Mpc scale allowed by the bispectrum, when the conformal coupling was broken at the beginning of inflation, with an optimal scenario where the conformal coupling is only broken close to the time at which the Mpc scale left the horizon. As also shown in \cite{Ferreira:2013sqa}, in this last case the constraints are significantly weaker. One can verify that the magnetic fields at the Mpc scale increase as $\rho_{\rm inf}$ decreases and it can be as strong as $10^{-15}$ G at $\rho_{\rm inf}^{1/4} \simeq 10^{-2}$ GeV in contrast to the standard scenario where a maximum for $B$ in terms of $\rho_{\rm inf}^{1/4}$ appears in this window of energy at $\rho_{\rm inf}^{1/4} \simeq 10^{3}$ GeV and with strength $\sim 10^{-27}$ G. For high scale inflation, $\rho_{\rm inf}^{1/4} \simeq 10^{15}$ GeV, both cases lead to maximal magnetic fields of strength $\sim 10^{-32}$ G.
In the same figure, we also compare the optimal scenario in this model with late breaking of the conformal coupling to the constraint in eq. (\ref{eq:bound}). The plot indicates that in this highly optimized scenario, the constraint on the magnetic fields is nearly $3$ orders of magnitude lower than the conservative upper bound.

%%%%%%%%%%%%%%%%%%%%%%%%%%%%%%%%%%%%%%%%%%%%%%%%%%%%%%%%%%%%%%%%%%%%%%%%
\section{Constraints from B-modes}

In the light of the very recent observation of tensor modes by BICEP2 \cite{Ade:2014xna} we will assume those results as correct, and derive the respective constraints.

The BICEP2 experiment quotes a tensor to scalar ratio $r=0.2^{+0.07}_{-0.05}$. In the simplest models of inflation where the graviton is the only component capable of generating the primordial tensor perturbations, the squared amplitude of tensor perturbations ($A_T$) is proportional to the energy scale of inflation
\beq
A_T^2=\frac{8}{ M_p^2} \lmk \frac{H}{2\pi} \rmk ^2.
\eeq 
On the other hand $r=A^2_T/A^2_R$ where $A_R^2=2.2\times10^{-9}$ is the amplitude of scalar perturbations. Therefore, it is straightforward to derive the energy scale of inflation to be,
\beq
H\simeq1.1\times 10^{14}\,  \txt{GeV}\quad \Ra  \quad \rho_{\rm inf} ^{1/4}\simeq2.2\times10^{16} \, \txt{GeV}.
\eeq

If the spectrum of gravitational waves is also nearly scale-invariant, as predicted, this result has deep consequences for inflationary magnetogenesis. 
The minimal value of $\al$ allowed for each constraint in eqs. (\ref{eq:BCal}, \ref{eq:PCal}, \ref{eq:fNLCal}) approaches the non-backreacting case: $\al=-2$. For these values of $H$ and $\al$ eq. (\ref{eq:B}) leads to a maximal magnetic field in $f^2(\phi)F_{\mu \nu}F^{\mu \nu}$ models of strength $B_k\simeq 8.1\times10^{-35} \lmk k/k_{\txt{Mpc}} \rmk$ G, which should be compared with the conservative upper bound, which is valid only for $k>\txt{Mpc}^{-1}$, yielding $B_k<1.3\times10^{-30}\lmk k/k_{\txt{Mpc}} \rmk^{5/4}$ G. As we can see in Fig. (\ref{fig:Btens}) the upper bound does not allow for $B\gtrsim10^{-30}$G at Mpc scale, although it does at smaller scales. In $f^2(\phi)F_{\mu \nu}F^{\mu \nu}$ models that value of the magnetic field is only allowed below the kpc scale.
 
\bef
\begin{centering}
\includegraphics[width=8.4cm,height=6cm]{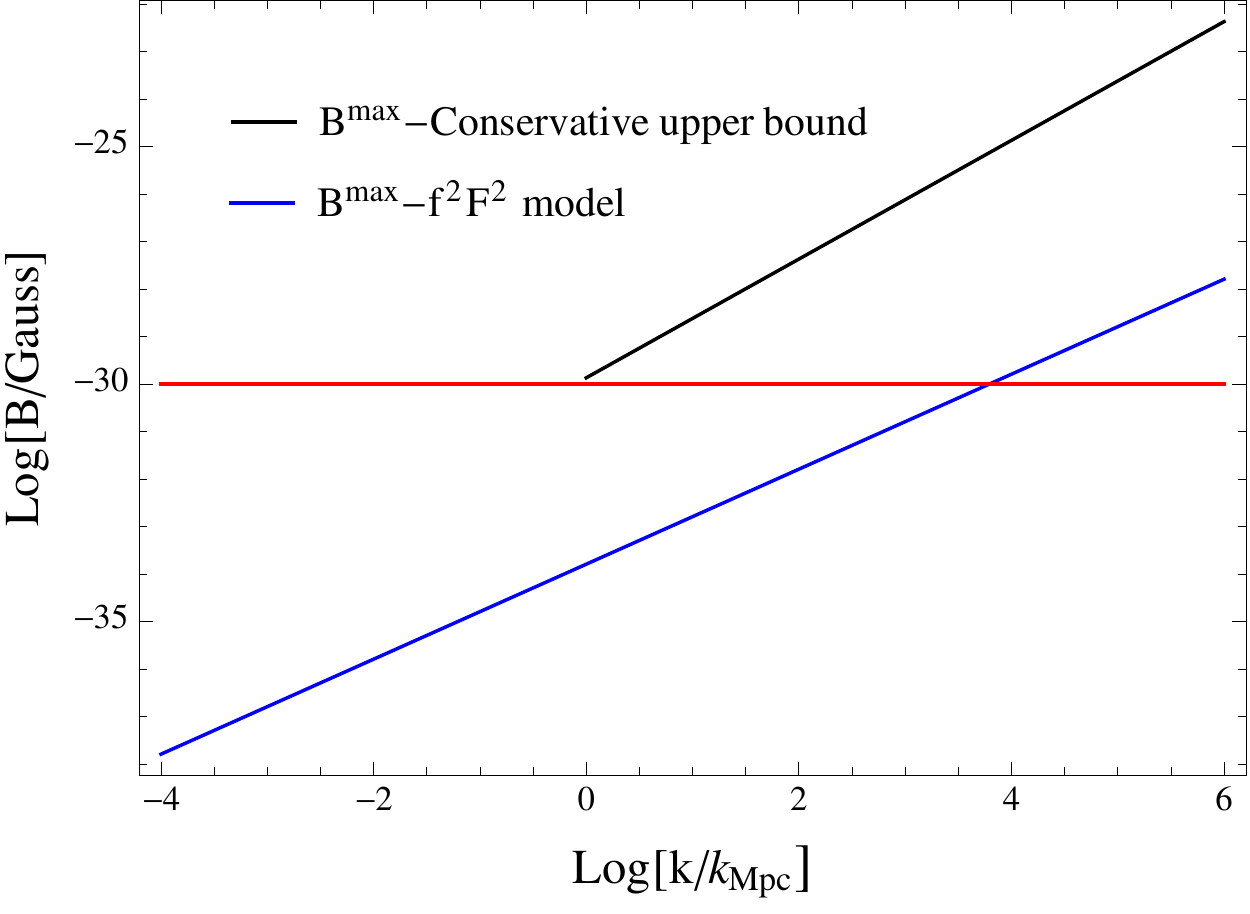}
\caption{Maximal magnetic field allowed by the conservative upper bound derived in \cite{Fujita:2014sna} and $f^2(\phi) F^2$ models from the horizon scale up to pc scale, taking into account the recent observations of tensor modes by BICEP2.}
\label{fig:Btens}
\end{centering}
\eef

%%%%%%%%%%%%%%%%%%%%%%%%%%%%%%%%%%%%%%%%%%%%%%%%%%%%%%%%%%%%%%%%%%%%%%%%
\section{Conclusions}

In this paper we have analyzed the constraints on the $f^2(\phi)F_{\mu \nu}F^{\mu \nu}$ model of inflationary magnetogenesis. We have focussed on constraints coming from the observed CMB anisotropies and B-modes.  We have found the maximal magnetic fields achievable in the $f^2(\phi)F_{\mu \nu}F^{\mu \nu}$ model, and compared with the recent constraint derived by Fujita and Yokoyama \cite{Fujita:2014sna}. We find that backreaction constraints provide stronger constraints than the constraints from anisotropies, if inflation lasts more than about $5$ e-folds than the minimal amount of e-foldings required to solve the horizon problem. However, if inflation is shorter, we find that the constraints from anisotropies become stronger. 

An important outcome of this analysis is the fact that, as was also pointed out in \cite{Ferreira:2013sqa}, strong magnetic fields are permissible only when the energy scale of inflation is significantly lowered. As we have shown here, in the standard scenario where the conformal coupling is broken at the beginning of inflation, the maximal magnetic field at the horizon scale increases as the energy scale of inflation decreases yielding $B \simeq 10^{-19}$ G at $\rho^{1/4}_{\rm inf} \simeq 10$ MeV and only $\simeq 10^{-35}$ G at $\rho^{1/4}_{\rm inf}\simeq10^{15}$ GeV. In the most optimal scenario, where inflation lasts close to the minimum amount of e-folds allowed and the conformal coupling is broken at the Mpc scale, the maximal magnetic fields allowed are $\sim 10^{-15}$ G at the Mpc scale for $\rho^{1/4}_{\rm inf}\simeq10$ MeV while for high scale inflation with $\rho_{\rm inf}^{1/4} \simeq 10^{15}$ GeV, the strength drops again to $\simeq10^{-32}$ G. In comparison with the upper bound derived in \cite{Fujita:2014sna}, our best case scenario is nearly 3 orders of magnitude lower. We have also studied similar constraints arising from the trispectrum on the magnetic field strength, but they are very similar to the bispectrum constraints.

The results presented here together with the previous work in the literature clearly indicate that it might be extremely difficult to explain the observations of void magnetic fields with an inflationary mechanism. This pessimism is even more justified in light of the recent observations of tensor modes by the BICEP2 experiment. If that observation is confirmed by other experiments, it would mean that the energy scale of inflation is $\rho^{1/4}_{\rm inf} \simeq 10^{16}$ GeV. With such a high scale inflation the upper bound allows for a maximal magnetic field of strength $B_k=1.3\times10^{-30}\lmk k/k_{\txt{Mpc}} \rmk^{5/4}$ G, for $k>\txt{Mpc}^{-1}$, while for $f^2(\phi)F_{\mu \nu}F^{\mu \nu}$ models the maximal value allowed for the magnetic field is $B_k=8.1\times10^{-35} \lmk k/k_{\txt{Mpc}} \rmk$ G. These results basically exclude the explanation of void magnetic fields by inflationary mechanisms, unless the BICEP2 B-mode signal is not due to primordial gravitational waves \cite{Bonvin:2014xia,Liu:2014mpa,Mortonson:2014bja} or gravitational waves are produced during inflation by some non-standard mechanism \cite{Mukohyama:2014gba,Cook:2011hg} keeping the scale of inflation low.

We conclude that these constraints are very stringent even for the generation of seed magnetic fields. Although an inflationary explanation for the seeds is still allowed by the upper bound, their generation in $f^2(\phi)F_{\mu \nu}F^{\mu \nu}$ models is no longer possible at scales larger than kpc.

\section*{Acknowledgments}

RKJ is supported by an individual postdoctoral fellowship from the Danish council for independent research in Natural Sciences. MSS is supported by a Jr. Group Leader Fellowship from the Lundbeck Foundation. The CP$^3$-Origins centre is partially funded by the Danish National Research Foundation, grant number DNRF90. 

%%%%%%%%%%%%%%%%%%%%%%%%%%%%%%%%%%%%%%%%%%%%%%%%%%%%%%%%%%%%%%%%%%%%%%%%

\bibliographystyle{JHEP}
\bibliography{AnisotropicConstraints}

%%%%%%%%%%%%%%%%%%%%%%%%%%%%%%%%%%%%%%%%%%%%%%%%%%%%%%%%%%%%%%%%%%%%%%%%
\end{document}